\documentstyle[11pt]{article}\textheight 230mm\textwidth 150mm
\newcommand{\be}{\begin{eqnarray}}
\newcommand{\eel}[1]{\label{#1}\end{eqnarray}}
\newcommand{\r}[1]{(\ref{e:#1})}
\newcommand{\vb}{{\cal h}}
\newcommand{\hb}{{\cal i}}
\newcommand{\ca}{{\cal C}}

\newcommand{\el}{{\em l}}
\newcommand{\baca}{\bar{\cal C}}
\newcommand{\lra}{{\leftrightarrow}}
\newcommand{\ra}{{\rightarrow}}
\newcommand{\Ra}{{\Rightarrow}}
\newcommand{\nn}{\nonumber}
\newcommand{\eg}{{\em e.g.\ }}
\newcommand{\ie}{{\em i.e.\ }}
\newcommand{\al}{\alpha}
\newcommand{\ga}{{\gamma}}
\newcommand{\ve}{{\varepsilon}}
\newcommand{\la}{{\lambda}}

\newcommand{\del}{{\delta}}

\newcommand{\Om}{\Omega}
\newcommand{\pet}{{\cal P}}
\newcommand{\bapet}{\bar{\pet}}
\newcommand{\dagg}{^{\dag}}
\newcommand{\bett}{{\bf 1}}
\newcommand{\halv}{\frac{1}{2}}
\begin{document}
\begin{titlepage}
\noindent
G\"{o}teborg ITP 94-32\\
December 1994\\
hep-th/9501004\\

\vspace*{5 mm}
\vspace*{20mm}
\begin{center}{\LARGE\bf Solving general gauge theories\\ on inner product
spaces.}
\end{center}
\vspace*{3 mm} \begin{center} \vspace*{3 mm}

\begin{center}Igor Batalin\footnote{Permanent address: Lebedev Physical
Institute,
1179234 Moscow, Russia} and Robert Marnelius \\ \vspace*{7 mm}
{\sl Institute of Theoretical Physics\\
Chalmers University of Technology\\
S-412 96  G\"{o}teborg, Sweden}\end{center}
\vspace*{10 mm}
\begin{abstract}
By means of a generalized quartet mechanism we show in a model independent way
that
a BRST quantization on an inner product space leads to physical states of the
form
$$|ph\hb=e^{[Q,
\psi]} |ph\hb_0$$ where $Q$ is the nilpotent BRST  operator,
$\psi$  a  hermitian fermionic gauge fixing operator, and $|ph\hb_0$   BRST
invariant states determined by a {\em hermitian} set of BRST doublets in
involution.
$|ph\hb_0$ does not belong to an inner product space although $|ph\hb$ does.
Since
the BRST quartets are split into two sets of hermitian BRST doublets there are
two choices for $|ph\hb_0$ and the corresponding $\psi$. When
applied to general, both irreducible and
reducible, gauge theories of arbitrary rank within the BFV formulation we find
that
 $|ph\hb_0$
are trivial BRST invariant states  which only depend on the matter
variables for one set of solutions, and
for the other set $|ph\hb_0$ are solutions of a Dirac quantization.
 This generalizes previous Lie group solutions obtained by means of a
bigrading.\end{abstract}\end{center}\end{titlepage}

\setcounter{page}{1}
\section{Introduction.}
In ref.\cite{Simple} simple expressions for the solutions of a BFV-BRST
quantization on
inner product spaces was obtained for arbitrary  irreducible Lie group
gauge theories
with finite number of degrees of freedom. More
precisely it was shown that provided one makes use of dynamical  Lagrange
multipliers and antighosts the physical states $|ph\hb$,
satisfying $Q|ph\hb=0$, may be written as
\be
&&|ph\hb=e^{[Q, \psi]}|ph\hb_0
\eel{e:1}
where $\psi$ is a specific hermitian fermionic gauge fixing operator
 with ghost number
minus  one, and
where $|ph\hb_0$ is a trivial BRST invariant state which only depends on the
matter
variables.
These are formal solutions and one has to associate certain quantum
prescriptions of
the involved operators in order for these formal solutions  to be true
nontrivial
solutions.  The basic general
quantization rule is that the unphysical degrees of freedom represented by
ghosts
and antighosts,  Lagrange multipliers and  gauge degrees of freedom are to
be quantized in an opposite manner, \ie one with positive and the other with
indefinite metric states so that they together form states built of half
positive and
half indefinite metric state spaces \cite{Proper}. (Further properties of these
solutions are given in \cite{Path} and \cite{MQ}.)

In \cite{Gauge} it was shown that $|ph\hb$ may also be related to solutions of
 a Dirac
 quantization. This
relation was also shown to be of the form \r{1} but where the ghost fixed
states
$|ph\hb_0$ are
 solutions of a
Dirac quantization, and where $\psi$ has to be chosen differently.

In this paper we give a general setting for solutions of the form \r{1} and
prove
that the results of \cite{Simple,Gauge} may be generalized to arbitrary,
 both  irreducible
and reducible  gauge
theories.

Throughout the paper we make use of supercommutators defined by
\be
&&[A, B]=AB-BA(-1)^{\varepsilon_A\varepsilon_B}
\eel{e:01}
where $\varepsilon_A$ and $\varepsilon_B$ are the Grassmann parities of the
operators
$A$ and $B$ respectively. ($\varepsilon_A=0$ for even $A$ and $\varepsilon_A=1$
for
odd $A$.)

The paper is organized as follows: In section 2 we review some general
properties of
BRST quantization on inner product spaces. In section 3 we prove formula \r{1}
in a
general setting and discuss some of its properties, and in section 4 we
illustrate
these properties in a simple  case corresponding to \eg abelian gauge
theories. In section 5 we start to look for more explicit realizations. We
consider
first the minimal sector of  general irreducible gauge theories of
arbitrary rank within the BFV scheme. In section
6 we treat the nonminimal sector with dynamical Lagrange multipliers and
antighosts. In
section 7 we apply the results of section 3 and give some  properties and
interpretations. In section 8 we extend the previous results to general
reducible gauge theories.
 Finally we conclude the paper in
section 9.

\section{BRST quantization on inner product spaces}
Consider a gauge theory with a conserved nilpotent BRST charge operator $Q$.
Let, furthermore, the associated state space $V$ be a nondegenerate inner
product
space. The states in $V$ may then be subdivided into singlets and doublets
under $Q$
as follows (\cite{KO,Nish}):
\be
1)&&|s\hb\in V \;\mbox{is a singlet if}\; Q|s\hb=0,\;\;\;|s\hb\neq
Q|u\hb\;\mbox{any}\;|u\hb\in V\nn\\ 2)&&|d\hb,\,|p\hb\in V
\;\mbox{is a doublet if}\;
|d\hb=Q|p\hb\neq0 \eel{e:101}
This  subdivision is not unique since it is invariant under
\be
&&|s\hb\;\ra\;|s\hb+|d'\hb\nn\\
&&|p\hb\;\ra\;|p\hb+|s'\hb+|d''\hb
\eel{e:102}
We may therefore impose further conditions. First we may choose
$|d'\hb$ and $|s'\hb$
in \r{102} so that
\be
&&\vb s|p\hb=0,\;\;\;\forall\;|s\hb,\;|p\hb
\eel{e:103}
is always valid \cite{Nish}. In this case $V$ is
 divided into a direct sum of singlet and
doublet states:
\be
&&V=V_S\oplus V_D
\eel{e:104}
The nondegeneracy of the inner product of $V$ forbid the existence of $|d\hb$'s
different from zero such that $\vb d|p\hb=0$ for all $|p\hb$ since this
condition is
equivalent to $\vb d|u\hb=0$ for all $|u\hb\in V$. Eq.\r{104} implies therefore
that
$V_S$ is a representation of the BRST cohomology $Ker Q/ Im Q$
($|s\hb,\;|d\hb\in Ker
Q,\;\;|d\hb\in Im Q$). We may furthermore choose a $|d''\hb$ to every $|p\hb$
in
\r{102} so that \cite{Nish}:
\be
&&\vb p|p'\hb=0
\eel{e:1020}
for all $|p\hb$-states. In the following we always require \r{103}
and often \r{1020}
as well.

One way to determine $V_S$ is through the Hodge decomposition implied by the
coBRST
charge \cite{Nish,Spi,RRy,KvH}. This construction requires the existence of an
even,
hermitian metric operator $\eta$ satisfying
\be
&&\vb u|\eta|u\hb\geq 0,\;\;\forall |u\hb\in V,\;\;\;\eta^2=\bett
\eel{e:105}
This means that $V$ is a bilinear form on a Hilbert space which is a natural
restriction. The coBRST charge is then defined in terms of $\eta$ through
\be
&&^*Q\equiv\eta Q\eta
\eel{e:106}
This definition implies that $^*Q$ is nilpotent. One may
now show that all $|p\hb$'s
may be chosen to have the form
\be
&&|p\hb=\,^*Q|u\hb
\eel{e:107}
which automatically satisfies \r{1020} while \r{103} requires
\be
&&^*Q|s\hb=0
\eel{e:108}
Since one may  show that
\be
&&\triangle |s\hb=0\;\Leftrightarrow\;Q|s\hb=\,^*Q|s\hb=0
\eel{e:109}
where $\triangle \equiv[Q,\, ^*Q]_+$, $V_S$ is also the space of BRST harmonic
states.
Eq.\r{104} is then the Hodge decomposition ($|u\hb=|s\hb+Q|u'\hb+\,^*Q|u''\hb$
any
$|u\hb\in V$).

There is no unique relation between the metric operator $\eta$ and the coBRST
charge
$^*Q$. We may \eg   factorized $\eta$ as follows
\be
&&\eta=\eta_S\eta_D=\eta_D\eta_S
\eel{e:110}
where
\be
&&[Q, \eta_S]=0,\;\;\;[Q, \eta_D]\neq 0
\eel{e:111}
This implies
\be
&&^*Q=\eta_DQ\eta_D
\eel{e:112}
which in turn implies that $\eta_D$ determines all the properties within the
coBRST
approach. Notice that $\eta_D$ must be nontrivial which means that the
unphysical
degrees of freedom must contain indefinite metric states. In fact, the
unphysical
degrees of freedom must be quantized with half  positive and half  indefinite
metric states since the nondegenerate doublet space $V_D$ is divided into two
equally
large subspaces  of zero norm states, \ie $Im Q$ and $Im ^*Q$. Notice that the
singlet states $|s\hb$ only have positive norms if $\eta_S=\bett$, which thus
is a
condition one has to impose on physical theories. It is essentially equivalent
to the
completeness condition of Spiegelglas \cite{Spi}: All zero norm states in $Ker
Q$ are
in $ImQ$. (However, this condition also allow for $\eta_S=-\bett$.)

We now turn to another way to determine $V_S$ in \r{104} which
has a less invariant form but which we expect to be related to the above
approach.
The starting point is the following argument:
 If we assume that the BRST doublets $|d\hb$ and
$|p\hb$ may be represented as follows (this depends on the basis of $V$)
 \be
&&|p_i\hb=C_i\dagg|u\hb,\;\;\;|d_i\hb=QC_i\dagg|u\hb\; \mbox{any}\; |u\hb\in V
\eel{e:113}
then condition \r{103} requires
\be
&&C_i|s\hb=0
\eel{e:114}
which in turn implies
\be
&&B_i|s\hb=0,\;\;\;B_i\equiv[Q, C_i]
\eel{e:1141}
The operator doublets, $D_r\equiv(C_i,B_i)$, must then satisfy the consistency
conditions
\be
&&[D_r, D_s]={K_{rs}}^tD_t
\eel{e:115}
which implies
\be
&&[D_r\dagg, D_s\dagg]=D_t\dagg K_{sr}^{\dag \;t}
\eel{e:116}
(That auxiliary conditions of the form \r{114}-\r{1141} may always be imposed
was
demonstrated in  \cite{Aux,Auxfor}.) The nondegeneracy of the inner product of
$V_D$
requires now \be
&&[D_r, D_s\dagg]\;\mbox{ is an invertible matrix operator}
\eel{e:117}
 even between singlet states. From \r{115} and \r{116} this requires that
the set $\{D_r\}$ is linearly independent of  $\{D_r\dagg\}$ and that they
together
constitute a set of (generalized) BRST quartets \cite{KO,Nish,Auxfor}. One may
notice
that if singlet states are assumed to be of the form
\be
&&|s_i\hb=A_i\dagg|0\hb
\eel{e:1160}
where $|0\hb$ is a singlet vacuum state then the singlet operators  must
satisfy
\be
&&[D_r, A_i\dagg]={a_{ir}}^sD_s\nn\\
&&[A_i, D_r\dagg]=D_s\dagg{a_{ir}^{\dag\; s}}
\eel{e:1161}
for consistency.

The property \r{1020} of the $|p\hb$-states requires at least that the
commutator
$[C_i, C_j\dagg]$
vanish between singlet states. This together with the property
\be
&&[B_i, B_j\dagg]=[Q, [C_i, B_j\dagg]],
\eel{e:1172}
which follows from the representation \r{1141} by means of the Jacobi
identities,
implies now that \r{117} requires
  \be
&&[C_i, B_j\dagg]\;\mbox{ is an invertible matrix operator}
\eel{e:1173}
(cf. \cite{KO,Nish}). Due to the definition \r{1141} of the $B$-operators this
condition requires that the set of $C$-operators $\{C_i\}$ is divided into two
equally
large sets, one with bosons and one with fermions, which in turn implies  that
the
index $i$ must run over an even number.

The above approach to a representation $V_S$ of the BRST cohomology requires
us to find
a maximal irreducible set of operator doublets $\{D_r\}$ satisfying \r{115}.
(We
shall call such a set a complete set of doublets.) $V_S$ is then determined by
conditions \r{114}-\r{1141} \ie \be
&&D_r|s\hb=0
\eel{e:118}
At least in the case when there is a ghost number operator $N$ satisfying
\be
&&[N, Q]=Q
\eel{e:1181}
one may prove that
\be
&&Q|s\hb=0
\eel{e:119} will always be implied by \r{118} since one then has
\be
&&Q=a^rD_r
\eel{e:120}

In the above approach we expect that one always can arrange the doublets so
that
there exists a nilpotent coBRST operator $^*Q$ satisfying
\be
&&C_i=[\,^*Q, B_i]
\eel{e:1170}
This implies by means of the Jacobi identities that
\be
&&[C_i, C_j\dagg]=[\,^*Q, [B_i, C_j\dagg]]
\eel{e:1200}
Thus, if the singlet states $|s\hb$ are coBRST invariant, \ie satisfies
\r{108}, then
$[C_i, C_j\dagg]$ vanishes between the singlet states. It follows also that
singlet
operators $A_i$ should satisfy \be
&&[Q, A_i]=[\,^*Q, A_i]=0
\eel{e:1171}
which implies
\be
&&[C_i, A_j]=[\,^*Q, [Q, [C_i, A_j]]]\nn\\
&&[B_i, A_j]=[Q, [\,^*Q, [B_i, A_j]]]
\eel{e:11721}
\ie these commutators vanish between singlet states. We expect that a coBRST
charge
defined by \r{1170} and satisfying nilpotency is equivalent to a coBRST
charge defined by \r{106}. Hence,  in this case
\r{118} should be equivalent to  $\triangle|s\hb=0$.

\setcounter{equation}{0}
\section{A general setting for formula (1.1).}

Let as before $Q$ be a nilpotent BRST charge operator defined on a
nondegenerate
inner product space $V$. Determine then a maximal elementary set of operator
doublets
$\{ D_{(1)r} \}$ which are in involution. However, in distinction to the
previous section we require now that $D_{(1)r} $ are {\em hermitian}. If we
then
determine singlet states by
\be
&&D_{(1)r} |s\hb_0=0
\eel{e:2}
it is clear that the solutions $|s\hb_0$ cannot belong to an inner product
space since
\r{117} is not satisfied. In this case it is, however, natural to expect that
there
exists an equally large set of BRST doublets $\{ D_{(2)r}\}$ whose elements
also are hermitian and
 in involution satisfying
\be
&&[D_{(1)r}, D_{(2)s}]\; \mbox{ is an invertible matrix operator}
\eel{e:3}
In fact, this is just another polarization of the unphysical operators. Due to
the
hermiticity of the doublets there involution relations satisfy
\be
&&[D_{(\l)r}, D_{(l)s}]={K_{(\l)rs}}^tD_{(\el)t}=
D_{(l)t}{K^{\dag\;\;t}_{(\l)sr}},\;\;\;\l=1,2
\eel{e:4}

We assume now that we have two such dual sets of hermitian BRST doublets
$D_{(1)r}$
and $D_{(2)r}$. It follows then that we may define two sets of
singlet states $|s\hb_0^{(1,2)}$ by
\be
&&D_{(\l)r}|s\hb_0^{(\l)}=0,\;\;\;\l=1,2
\eel{e:5}
neither of which belong to an inner product space but whose bilinear form
$\,^{(1)}_0\vb s|s\hb_0^{(2)}$ might be finite.

We shall now prove that the singlet states $|s\hb_0^{(\el)}$ may be related to
singlet states $|s\hb^{(\el)}$ on inner product spaces under certain
conditions.
This relation  is given by
\be
&|s\hb^{(\el)}=e^{[Q, \psi_{\el}]}|s\hb_0^{(\el)}
\eel{e:6}
where $\psi_{\el}$ is a specific odd hermitian operator. Obviously \r{5} imply
\be
&&D'_{(\el)r}|s\hb^{(\el)}=0,\;\;\;\el=1,2
\eel{e:7}
where
\be
&&D'_{(\el)r}\equiv e^{[Q, \psi_{\el}]}D_{(\el)r}e^{-[Q, \psi_{\el}]}
\eel{e:8}
also are BRST doublets since $[Q, \psi_{\el}]$ is BRST invariant. This relation
imply
\be
&&{D'}\dagg_{(\el)r}=e^{-[Q, \psi_{\el}]}D_{(\el)r}e^{[Q,
\psi_{\el}]}\neq D'_{(\el)r}
\eel{e:9}
which provides for the possibility to satisfy \r{117}. A necessary condition
for this
is that $C_{(\el) i}$ and $B_{(\el) i}$ of $D_{(\el)r}\equiv(C_{(\el) i},
B_{(\el) i})$
each consists of half bosons and half fermions.
This will therefore be assumed to be the case in the following.
 We propose then that $\psi_{\el}$ in \r{6} should be expressed in
terms of all the $C$-operators in the dual set of doublets according to the
formula
\be
&&\psi_{1}\equiv  C_{(2)a}^{(b)}C_{(2)}^{(f)a}\nn\\
&&\psi_{2}\equiv  C_{(1)a}^{(b)}C_{(1)}^{(f)a}
\eel{e:10}
where $C_{(\el)a}^{(b)}$ and $C_{(\el)a}^{(f)}$ are the bosonic and fermionic
operators of $C_{(\el)i}$($\equiv(C_{(\el)a}^{(b)},C_{(\el)a}^{(f)})$) and
where the
indices $a$, $b$ are supposed to be raised and lowered by means of a constant
symmetric metric. Eq. \r{10} is a hermitian expression if $C_{(\el)a}^{(b)}$
and
$C_{(\el)a}^{(f)}$ commute which we assume (otherwise we have to symmetrize
\r{10}).
Eq. \r{10} implies \be
&&[Q, \psi_{1}]= C_{(2)a}^{(b)} B_{(2)}^{(b)a}-i
B_{(2)a}^{(f)}C_{(2)a}^{(f)a}\nn\\
&&[Q, \psi_{2}]= C_{(1)a}^{(b)} B_{(1)}^{(b)a}-i
B_{(1)a}^{(f)}C_{(1)a}^{(f)a}
\eel{e:11}
where we have introduced hermitian $B$-operators defined by
\be
&&B_{(l)a}^{(f)}\equiv i[Q, C_{(l)a}^{(b)}],\;\;\;B_{(l)a}^{(b)}\equiv [Q,
C_{(l)a}^{(f)}]
\eel{e:1111}
Thus, the right-hand side of $[Q, \psi_{l}]$ involves the complete dual set of
doublets.

We assume now that the commutator $[C_{(1)i}, C_{(2)j}]$ vanish between the
bilinear form of the states $|s\hb^{(1)}$ and $|s\hb^{(2)}$ (cf. the statement
over
\r{1172} in the nonhermitian case given in the previous section) which means
that
\r{3} requires
\be
&&[B_{(1)i}, C_{(2)j}]\; \mbox{and}\;[B_{(2)i}, C_{(1)j}]\; \mbox{are
invertible
matrices}
\eel{e:12}
which due to the split into fermionic and bosonic operators in turn requires
\be
&&[B_{(1)a}^{(b)}, C_{(2)b}^{(b)}],\;[B_{(1)aa}^{(b)},
C_{(2)b}^{(b)}],\;[B_{(1)a}^{(f)}, C_{(2)b}^{(f)}],\; \mbox{and}\nn\\
&&[B_{(2)a}^{(f)},
 C_{(1)b}^{(f)}]\;
\mbox{are invertible matrix operators}
\eel{e:1211}
The BRST doublets fall then into quartets. These properties imply now
\be
&&[[Q, \psi_{(1)}], D_{(1)r}]={A_{r}}^{s}D_{(2)s}=-D_{(2)s}
{A^{\dag\; s}_{r}}\nn\\
&&[[Q, \psi_{(2)}], D_{(2)r}]={B_{r}}^{s}D_{(1)s}=-D_{(1)s}
{B^{\dag\;s}_{r}}
\eel{e:13}
where ${A_{r}}^s$ and ${B_{r}}^s$ are invertible matrix operators. It follows
then
that \be
&&D'_{(1)r}=e^{[Q, \psi_{1}]}D_{(1)r}e^{-[Q,
\psi_{1}]}=D_{(1)r}+{A_{r}}^{s}D_{(2)s}+O(D_{(2)}^2)\nn\\
 &&D'_{(2)r}=e^{[Q,
\psi_{2}]}D_{(2)r}e^{-[Q, \psi_{2}]}=D_{(2)r}+{B_{r}}^{s}D_{(1)s}
+O(D_{(1)}^2)
\eel{e:14}
and
\be
&&{D'}\dagg_{(1)r}=e^{[Q, \psi_{1}]}D_{(1)r}e^{-[Q,
\psi_{1}]}=D_{(1)r}-{A_{r}}^sD_{(2)s}+O(D_{(2)}^2)\nn\\
 &&{D'}\dagg_{(2)r}=e^{[Q,
\psi_{2}]}D_{(2)r}e^{-[Q, \psi_{2}]}=D_{(2)r}-{B_{r}}^sD_{(1)s}+O(D_{(1)}^2)
\eel{e:15}
where $O(D_{(\el)}^2)$ denotes nonlinear terms in the $D_{(\el)}$'s.
The fact that the
matrix operators ${A_{r}}^s$ and ${B_{r}}^s$ are nonsingular implies that
$\{D'_{(1)r}\}, \;\{(D'_{(1)r})^{\dag}\}$ as well as $\{D'_{(2)r}\}
\;\{(D'_{(2)r})^{\dag}\}$ are algebraically independent sets and constitute
generalized
BRST quartets, \ie the matrix operators $[D'_{(1)r}, (D'_{(1)s})\dagg]$ as well
as
$[D'_{(2)r}, (D'_{(2)s})\dagg]$  are nonsingular.
Since $\{D_{(1)r}\}$ and $\{D_{(2)r}\}$ are complete sets  the BRST doublets
$\{D'_{(1)r}\}$ and $\{D'_{(2)r}\}$ are also complete sets. This implies that
the
physical states $|s\hb^{(1,2)}$ in \r{6}
 each belongs to a nondegenerate physical state space
representing the BRST cohomology. The assertion about \r{6} is then proved.

\setcounter{equation}{0}
\section{A simple  example.}

In order to get some insight into the above general structure of general BRST
quantization we consider the simplest possible example when the hermitian
BRST doublets
$(C_{(\el) a}^{(b)}, C_{(\el)
a}^{(f)}, B_{(\el) a}^{(b)}, B_{(\el) a}^{(f)})$ are completely elementary, \ie
the
case when the only nonzero commutators are given by
\be
&&[C_{(1)a}^{(b)}, B_{(2)}^{(b)b}]=i\del^a_b=[C_{(2)
a}^{(b)}, B_{(1)
}^{(b)b}]\nn\\
&&[C_{(1)
a}^{(f)}, B_{(2)
}^{(f)b}]=\del^a_b=[C_{(2)
a}^{(f)}, B_{(1)
}^{(f)b}]
\eel{e:16}
In this case the coBRST charge operator is given by
\be
&&^*Q=C_{(1)a}^{(b)}C_{(2)}^{(f)a}+C_{(2)a}^{(b)}C_{(1)}^{(f)a}
\eel{e:17}
which implies
\be
&&[^*Q,B_{(\el)a}^{(b)}]=iC_{(\el)a}^{(f)}\nn\\
&&[^*Q,B_{(\el)a}^{(f)}]=C_{(\el)a}^{(b)}
\eel{e:18}
The BRST charge itself is of the form
\be
&&Q=B_{(1)a}^{(b)}B_{(2)}^{(f)a}+B_{(2)a}^{(b)}B_{(1)}^{(f)a}
\eel{e:19}
with the properties
\be
&&[Q,C_{(\el)a}^{(b)}]=-iB_{(\el)a}^{(f)}\nn\\
&&[Q,C_{(\el)a}^{(f)}]=B_{(\el)a}^{(b)}
\eel{e:20}
Notice that both $Q$ and $^*Q$ are hermitian.

Consider now the antihermitian operator
\be
&&R\equiv -iB_{(1)}^{(b)a}C_{(2)a}^{(b)}-iB_{(2)}^{(b)a}C_{(1)a}^{(b)}
+B_{(1)}^{(f)a}C_{(2)a}^{(f)}+B_{(2)}^{(f)a}C_{(1)a}^{(f)}
\eel{e:21}
It satisfies
\be
&&[R, B^a]=B^a,\;\;\;[R, C_a]=-C_a
\eel{e:22}
which implies
\be
&&[R, Q]=2Q,\;\;\;[R, \,^*Q]=-2\,^*Q
\eel{e:2211}
R may be split into two pieces
\be
&&R=R_1+R_2
\eel{e:23}
such that if the conjugate pair ($B_{(2)}^{(b)a},\;C_{(1)a}^{(b)}$)
is in $R_1$, then
the conjugate pair ($B_{(1)}^{(f)a},\;C_{(2)a}^{(f)}$) is in $R_2$ or vice
versa.
Then
$R_i$ satisfies
\be
&&[R_i, Q]=Q,\;\;\;[R_i, \,^*Q]=-\,^*Q\;\;\;\;i=1,2
\eel{e:24}
Thus, $R_1$ (or $R_2$) properly chosen  may be identified with the ghost number
operator $N$ in \r{1181} if we have no operators with larger ghost number than
$\pm
1$. This  implies then that half of the BRST doublets are ghosts. (If there are
operators with larger ghost number, then more than half of the doublets have
nonzero
ghost number.)

Consider now the BRST laplacian $\triangle=[^*Q, Q]$. It is here given by
\be
&&\triangle=B_{(1)}^{(b)a}C_{(2)a}^{(b)}+C_{(1)}^{(b)a}B_{(2)a}^{(b)}
-iB_{(1)}^{(f)a}C_{(2)a}^{(f)}+iC_{(1)}^{(f)a}B_{(2)a}^{(f)}=\nn\\
&&=B_{(2)}^{(b)a}C_{(1)a}^{(b)}+C_{(2)}^{(b)a}B_{(1)a}^{(b)}
-iB_{(2)}^{(f)a}C_{(1)a}^{(f)}+iC_{(2)}^{(f)a}B_{(1)a}^{(f)}
\eel{e:25}
It is clear that
\be
&&\triangle|s\hb_0=0
\eel{e:26}
implies either
\be
&&D_{(1)r}|s\hb_0=0,\mbox{  or  }D_{(2)r}|s\hb_0=0
\eel{e:27}
(One has to choose the original state space such that either of these
possibilities
allow for solutions.) These two possibilities are also equally well implied by
\be
&&Q|s\hb_0=\,^*Q|s\hb_0=0
\eel{e:28}

The odd hermitian operator $\psi_l$ in the formula \r{6} may now be chosen to
be
\be
&&\psi_l=[\,^*Q, F_l]
\eel{e:29}
where
\be
&&F_1=B_{(2)}^{(f)a}C_{(2)a}^{(f)},\;\;\;F_2=B_{(1)}^{(f)a}C_{(1)a}^{(f)}
\eel{e:30}
These expressions are antihermitian and reproduce \r{10} and satisfy
\be
&&[\triangle, F_l]=0
\eel{e:31}
which   is equivalent to
\be
&&[Q, [\,^*Q, F_l]]=[\,^*Q, [Q, F_l]]
\eel{e:32}
and implies for the inner product states in \r{6} that they also satisfies
\r{28} \ie
\be
&&Q|s\hb^{(\el)}=\,^*Q|s\hb^{(\el)}=0,\;\;\;l=1,2
\eel{e:33}
The nonhermitian doublets defined by \r{8} become here
\be
&&{C'}_{(1)a}^{(b)}={C}_{(1)a}^{(b)}-i{C}_{(2)a}^{(b)},\;\;\;
{C'}_{(1)a}^{(f)}={C}_{(1)a}^{(f)}+i{C}_{(2)a}^{(f)}\nn\\
&&{B'}_{(1)a}^{(b)}={B}_{(1)a}^{(b)}-i{B}_{(2)a}^{(b)},\;\;\;
{B'}_{(1)a}^{(f)}={B}_{(1)a}^{(f)}+i{B}_{(2)a}^{(f)}
\eel{e:34}
and
\be
&&{C'}_{(2)a}^{(b)}=-i({C'}_{(1)a}^{(b)})\dagg,\;\;\;
{C'}_{(2)a}^{(f)}=i({C'}_{(1)a}^{(f)})\dagg\nn\\
&&{B'}_{(2)a}^{(b)}=i({B'}_{(1)a}^{(b)})\dagg,\;\;\;
{B'}_{(2)a}^{(f)}=-i({B'}_{(1)a}^{(f)})\dagg
\eel{e:35}
Thus, in this case $\{D'_{(1)r},\;(D'_{(1)r})^{\dag}\}$ as well as
$\{D'_{(2)r},
\;(D'_{(2)r})^{\dag}\}$
are exactly the same sets. The BRST charge and the coBRST charge have the
following
form in terms of these nonhermitian doublets
\be
&&Q=\frac{i}{2}\left(B^{\dag(b)}_aB^{(f)a}-B^{\dag(f)}_aB^{(b)a}\right)\nn\\
&&^*Q=\frac{i}{2}\left(C^{\dag(f)}_aC^{(b)a}-C^{\dag(b)}_aC^{(f)a}\right)
\eel{e:36}
where we have made use of the short-hand notation
\be
&&{B}_{a}^{(b,f)}\equiv {B'}_{(1)a}^{(b,f)},\;\;\;
{C}_{a}^{(b,f)}\equiv {C'}_{(1)a}^{(b,f)}
\eel{e:37}

We are now going to apply the above general properties to the BFV formulation
of
general gauge theories.

\setcounter{equation}{0}
\section{Gauge theories of arbitrary rank. The minimal sector.}
 Consider a classical theory whose Hamiltonian formulation is
defined on a phase space
$\Gamma$ of
dimension $2n$. It contains  $m\leq n$ algebraically  independent first class
constraints
\be
&&\theta_a=0
\eel{e:201}
 The constraint variables $\theta_a$ are
assumed to be real with Grassmann parity $\varepsilon_a (=0,1)$. Their  algebra
in
terms of  the  Poisson bracket on $\Gamma$ is   of the form
 \be
 &&\{\theta_a, \theta_b\}=U_{ab}^{\;\;\;c}\theta_c
\eel{e:202}
where $U_{ab}^{\;\;\;c}$ are arbitrary real functions on
$\Gamma$
consistent with the Jacobi identities. After  quantization
$\theta_a$ and $U_{ab}^{\;\;\;c}$  are turned into hermitian operators
satisfying a
 commutator algebra of the form
\be
&&[\theta_a, \theta_b]=i\halv\left(\theta_cU_{ab}^{\;\;\;c}+U_{ab}^{\;\;\;c}
\theta_c\right) + \ldots
\eel{e:203}
where
the precise form of the right-hand side depends on the quantization
prescriptions for
$\theta_a$ and $U_{ab}^{\;\;\;c}$. Anyway, for whatever choice  made it is
obvious that the Dirac quantization
\be &&\theta_a|\;\hb=0
\eel{e:204}
is not consistent if $U_{ab}^{\;\;\;c}$ are nontrivial operators which does not
commute with
$\theta_a$ in \r{203}. As Dirac writes on page 70 in \cite{Dirac} "when we go
over to
the quantum
theory we must insist that the coefficients [$U_{ab}^{\;\;\;c}$] are on the
left" in
\r{203}. A precise solution of this dilemma is provided by the
 BRST quantization.

We introduce therefore hermitian generalized Faddeev-Popov
ghosts $\ca^a$ and their conjugate momenta $\pet_a$ with Grassmann parity
$\varepsilon_a
+ 1$ satisfying \be
&&[\ca^a, \pet_b]=i\del^a_b,\;\;\;\pet_a\dagg=-(-1)^{\varepsilon_a}\pet_a
\eel{e:205}
 The BFV-BRST charge operator for the above model is then given
by \cite{FF,BF,BF1,BF2}
\be
 &&\Om=\sum_{i=0}^N\Om_i
 \eel{e:206}
where
\be
&&\Om_0\equiv\ca^a\theta_a,\;\;\;\Om_i\equiv
\Om_{a_1\cdots a_{i+1}}^{b_1\cdots
b_i}(\ca^{a_1}\cdots\ca^{a_{i+1}}\pet_{b_1}\cdots
\pet_{b_i})_{Weyl},\;\;\;i=1,\ldots,N
\eel{e:207} where "Weyl" indicates that the ghosts are Weyl ordered. $N$ may be
infinite provided the infinite sum in \r{206} makes sense. $\Om$ is  required
to
be hermitian which in turn requires  $\Om_i$ to be hermitian as well. According
to
ref.\cite{BF2} there always exists nilpotent hermitian expressions of the form
\r{206} and these solutions determine the precise form of the algebra \r{203}.

Now a hermitian and nilpotent $\Om$ may also be written in a  $\ca\pet$-ordered
form \cite{BF,BF1,BF2}
 \be
 &&\Om=\sum_{i=0}^N\Om'_i,\;\;\;\Om'_0\equiv\ca^a\theta'_a,\nn\\
&&\Om'_i\equiv {\Om'}_{a_1\cdots a_{i+1}}^{b_1\cdots
b_i}\ca^{a_1}\cdots\ca^{a_{i+1}}\pet_{b_1}\cdots\pet_{b_i},\;\;\;i=1,\ldots,N
 \eel{e:208}
and in this case nilpotency requires the algebra
\be
&&[\theta'_a, \theta'_b]=i{U'}_{ab}^{\;\;\;c}\theta'_c
\eel{e:209}
which always allow for a consistent Dirac quantization given by
\be
&&\theta'_a|\;\hb=0
\eel{e:210}
The structure functions
${U'}_{ab}^{\;\;\;c}$ in \r{209} are given by
\be
&&{U'}_{ab}^{\;\;\;c}=2(-1)^{\varepsilon_a}{\Om'}_{ab}^{\;\;\;c}
\eel{e:211}
Identifying the expressions \r{206} and \r{208}  one may always express
$\theta_a$,
$\Om_{a_1\cdots a_{i+1}}^{b_1\cdots
b_i}$ in terms of $\theta'_a$,
${\Om'}_{a_1\cdots a_{i+1}}^{b_1\cdots
b_i}$ or vice versa. One will then find relations of the type
\be
&&\theta'_a=\theta_a+\sum_{k=1}^Ni^k
\Om_{ab_1\cdots b_k}^{\;\;\;b_1\cdots b_k}
\eel{e:212}
 \be
&&{\Om'}_{ab}^{\;\;\;c}=\Om_{ab}^{\;\;\;c}+\sum_{k=1}^Ni^k
\Om_{abb_1\cdots b_k}^{\;\;\;
cb_1\cdots b_k}
\eel{e:213}
which implies that $\theta'_a$ in general are not hermitian.  Exceptions are
\eg
 gauge theories invariant under unimodular Lie
groups.

We may also choose a $\pet\ca$-ordered $\Om$:
\be
&&\Om=\sum_{i=0}^N\Om''_i,\;\;\;\Om''_0\equiv\theta''_a\ca^a,\nn\\
&&\Om''_i\equiv {\Om''}_{a_1\cdots a_{i+1}}^{b_1\cdots
b_i}\pet_{b_1}\cdots\pet_{b_i}\ca^{a_1}\cdots\ca^{a_{i+1}},
\;\;\;i=1,\ldots,N
\eel{e:2131}
and in this case $\Om^2=0$ yields
\be
&&[\theta''_a, \theta''_b]=i\theta''_c{U''}_{ab}^{\;\;\;c}
\eel{e:214}
where
\be
&&\theta''_a=(\theta'_a)\dagg,\;\;\;{U''}_{ab}^{\;\;\;c}=
(-1)^{\varepsilon_a\varepsilon_b}({U'}_{ab}^{\;\;\;c})\dagg
\eel{e:215}
These $\ca\pet$- and $\pet\ca$-ordered forms of $\Om$ will be used in the
following.

We  look now for simple solutions of the BRST condition
\be
&&\Om|ph\hb=0
\eel{e:216}
relaxing for the moment the condition that $|ph\hb$ should
belong to an inner product
space. We  shall then make use
of consistent auxiliary conditions \cite{Aux,Auxfor} eventually
expressed in terms of
hermitian BRST doublets in involution. Following \cite{principle}  one may
\eg look for solutions which have no ghost dependence\footnote{In a consistent
gauge
theory on inner product spaces the physical states should not contain any ghost
excitations in order to have positive norms}. This may be done by means of a
ghost
fixing of the form \cite{Aux} \be &&g_a|ph\hb=0,\;\;\;a=1,\dots,n \eel{e:217}
where $g_a$ are $m$ independent linear expressions of the ghosts
$\ca^a$ and $\pet_a$.
Consistency requires here that \r{217} must be accompanied by \be
&&[\Om, g_a]|ph\hb=0,\;\;\;a=1,\ldots,m
\eel{e:218}
and that $g_a$ and $[\Om, g_a]$ must satisfy a closed algebra
with all coefficients to the left.
 The latter requires that $\Om$ has a specific form except for the following
two
cases: \be
&(1)\;g_a=\ca^a,&(2)\;\;g_a=\pet_a
\eel{e:219}
In case (1) we have
\be
&&[\Om, \ca^a]=\sum_{i=0}^N {\Om''}_{a_1\cdots a_{i+1}}^{b_1\cdots
b_i}[\pet_{b_1}\cdots\pet_{b_i},
\ca^a]\ca^{a_1}\cdots\ca^{a_{i+1}}(-1)^{s_i},\nn\\
&&s_i\equiv(\varepsilon_a+1)(i+1
+\sum_{k=1}^{i+1}\varepsilon_{a_k})
\eel{e:220}
where the $\pet\ca$-ordered $\Om$ \r{214}
is used. In this case $\ca^a$ and $[\Om, \ca^a]$ trivially satisfy a closed
algebra.
(Notice that $[\Om, \ca^a]$ satisfy a closed algebra among themselves only for
theories of rank $0,1$.)

In case (2) we have
\be
&&[\Om, \pet_a]=\sum_{i=0}^N {\Om'}^{a_1\cdots a_{i}}_{b_1\cdots
b_{i+1}}[\ca^{b_1}\cdots\ca^{b_{i+1}}, \pet_a]\pet_{a_1}
\cdots\pet_{a_{i}}(-1)^{s_i}
\eel{e:221}
where we use the more suitable  $\ca\pet$-ordered $\Om$ \r{208}.
Also here do $\pet_a$
and $[\Om, \pet_a]$ satisfy a closed algebra without any restrictions on $\Om$.
The
physical states may either be chosen to satisfy
\be
&&\ca^a|ph\hb=0
\eel{e:222}
or
\be
&&\pet_a|ph\hb=0,\;\;\;[\Om, \pet_a]|ph\hb=\theta'_a|ph\hb=0
\eel{e:223}
Notice that these conditions by themselves automatically imply  the BRST
condition
\r{216} since \be
&&\Om=A_a\ca^a=\ca^a\theta'_a+K^a\pet_a
\eel{e:224}
Thus, in case (1) we are led to the trivial ghost fixed solutions of \r{222}
and in
case (2) we are led to the consistent Dirac quantization \r{223}.

A crucial question is now whether or not the constraint operators in \r{222}
and
\r{223} belong to BRST doublets. $\pet_a$ and $[\Om, \pet_a]$ are BRST doublets
provided $[\Om, \pet_a]$
represent $m$ algebraic independent operators, which is the case if we have an
irreducible gauge theory. In a reducible gauge theory there are linear
combinations of
$\pet_a$ which are genuine physical operators which means that there
are more genuine
physical states than those determined by \r{223}. Only
irreducible gauge theories are considered here. The   reducible case will be
considered in section 8.

The conditions \r{222} are obtained from
BRST doublets provided there exist $m$ independent hermitian operators
$\chi^a$ with
ghost number zero satisfying
\be
&&[\Om, \chi^a]=iM^a_{\;b}\ca^b
\eel{e:225}
where $M^a_{\;b}$ is a nonsingular matrix operator in the sense that
$[\Om, \chi^a]|ph\hb=0$  imply \r{222}. In this case
\be
&&\chi^a|ph\hb=0
\eel{e:226}
will imply \r{222}, and \r{222} will allow for \r{226} \cite{Aux}. Consistency
requires $\chi^a$ and $[Q, \chi^a]$ to satisfy a closed algebra with
all coefficients
to the left.
Since $\chi^a$ are naturally chosen to be independent of $\ca^a$ this implies
that they
must be in involution:
\be
&&[\chi^a, \chi^b]=V^{ab}_c\chi^c
\eel{e:2261}
$\chi^a$ represent unphysical degrees of freedom and they may be viewed
as  gauge fixing operators to the gauge generators $[\Om, \pet_a]$. Notice that
\be
&&[\chi^b, [\Om, \pet_a]]=(-1)^{\varepsilon_b}{M'}^b_{\;a}
\eel{e:227}
where ${M'}^b_{\;a}$ is given by
\be
&&{M'}^b_{\;a}={M}^b_{\;a}+(-1)^{\varepsilon_a\varepsilon_c+
\varepsilon_a+\varepsilon_c}i[{M}^b_{\;c},\pet_a]\ca^c+[\Om,[\chi^b, \pet_a]]
\eel{e:228}
which is obtained from the Jacobi identities. (The last term is zero if
$\chi^a$ is
independent of $\ca^a$.) One may easily convince oneself that \r{226} yields
\r{222} even if
$M^a_{\;b}$ in  \r{225}
is replaced by ${M'}^a_{\;b}$.

If there is no gauge fixing operator $\chi^a$ satisfying the above conditions
then
there exist genuine physical operators (not belonging to any BRST doublets)
with
positive ghost number, which in turn means that there are other physical states
than
those determined by \r{222}. Here we shall always assume that there exist
gauge fixing operators $\chi^a$ satisfying the above conditions and that the
gauge
theory is irreducible. In this case the solutions of
\be
&&\ca^a|ph\hb=\chi^a|ph\hb=0
\eel{e:229}
or
\be
&&\pet_a|ph\hb'=[Q, \pet_a]|ph\hb'=0
\eel{e:230}
each represent the genuine physical degrees of freedom, \ie they are singlet
states.
In fact we have arrived at the general setting given in section 3: The
conditions
\r{229}-\r{230} may be  expressed in terms of the following two dual sets of
{\em
hermitian} BRST doublets
\be
&&D_{(1)r}\equiv\{\chi^a, i^{\varepsilon_a+1}[\Om,
\chi^a]\},\;\;\;D_{(2)r}\equiv
\{i^{\varepsilon_a+1}\pet_a, i[\Om,
\pet_a]\}
\eel{e:231}
For these we require that $[D_{(1)r}, D_{(2)s}]$ is an invertible matrix
operator.
Following section 3 we require also that the $C$-operators $\pet_a$ and
$\chi^a$
essentially commute so that this  implies
\be
&&[\chi^a, [\Om, \pet_b]]\mbox{ and }[[\Om, \chi^a],  \pet_b]\nn\\
&&\mbox{ are invertible
matrix operators} \eel{e:232}
Indeed this condition is satisfied since we have already shown that
${M'}^a_{\;b}$ in
\r{228} is invertible when $\chi^a$ satisfies the doublet condition \r{225},
\ie
when $M^a_{\;b}$ in \r{225} is  invertible.

\section{The nonminimal sector.}
\setcounter{equation}{0}

Since our goal is a BRST quantization on inner product spaces
where the genuine physical states have ghost number zero in a consistent theory
\cite{Na}, we cannot in general make use of the solutions of \r{222} or \r{223}
in
expressions like \r{1} in the introduction since these solutions have ghost
number
$m/2$ and $-m/2$ respectively. In order for such solutions
always to have ghost number
zero we need to introduce dynamical Lagrange multipliers and antighosts into
the
theory. The resulting extended BFV-BRST  charge is then given by\footnote{Our
notations are in accordance with those of [10-12], [14] 
 except
for the interchange $\pet\lra\bapet$}.
 \be
&&Q=\Om+\bapet^a\pi_a
\eel{e:301}
where $\bapet^a$  are conjugate momenta to the $m$ antighosts $\baca_a$, and
where
$\pi_a$ are conjugate momenta to the $m$ Lagrange multipliers $\la^a$. Their
hermiticity properties and Grassmann parities are
\be
&&\bapet^{a\dag}=\bapet^a,\;\;\;{\baca_a}\dagg=-(-1)^{\varepsilon_a}\baca_a,
\;\;\;\pi_a\dagg=\pi_a,
\;\;\;{\la^a}\dagg=(-1)^{\varepsilon_a}\la^a\nn\\
&&\varepsilon(\bapet_a)=\varepsilon(\baca^a)=\varepsilon_a+1,\;\;\;
\varepsilon(\pi_a)=\varepsilon(\la^a)=\varepsilon_a
\eel{3011}
and they satisfy the commutation relations (the
nonzero part)
\be
&&|\baca_a, \bapet^b]=i\del^b_a,\;\;\;[\la^a, \pi_b]=i\del^a_b
 \eel{e:302}
Notice that only the ghosts and antighosts carry nonzero ghost numbers:
\be
&&gh(\ca)=-gh(\pet)=gh(\bapet)=-gh(\baca)=1
\eel{3021}

In this extended case we need further conditions in order to fix the ghost
dependence
of the physical states. The appropriate generalization of cases (1) and (2) in
the
previous section is then
 \be
&&(1)\;\;\;\ca^a|ph\hb_{(1)}=\baca_a|ph\hb_{(1)}=0,
\;\;\;\pi_a|ph\hb_{(1)}=0
 \eel{e:303}
\be
&&(2)\;\;\;\pet_a|ph\hb_{(2)}=\bapet^a|ph\hb_{(2)}=0,\;\;\;
[Q, \pet_a]|ph\hb_{(2)}=0
\eel{e:304}
In fact these conditions are the only consistent extensions of the previous
cases
which make the physical states have ghost number zero (such conditions were
considered in \cite{SH,Aux}).
Notice that \r{303} and \r{304} are sufficient to make $|ph\hb_{(1,2)}$ BRST
invariant. However, as before they do not completely fix the physical states to
a
representation of the genuine physical degrees of freedom. In order to reduce
$|ph\hb_{(1,2)}$ to singlet states and comply with section 3 we have to
impose the gauge fixing conditions
\be
&&\chi^a|ph\hb_{(1)}=0,\;\;\;\Lambda^a|ph\hb_{(2)}=0
\eel{e:305}
where the hermitian gauge fixing operators $\chi^a$ and $\Lambda^a$ have ghost
number
zero and Grassmann parity $\varepsilon_a$, and are required to satisfy the
conditions
 \be
&&[Q, \chi^a]|ph\hb_{(1)}=0\;\; \Ra \;\;\ca^a|ph\hb_{(1)}=0\nn \\
&&[Q, \Lambda^a]|ph\hb_{(2)}=0\;\;\Ra  \;\;\bapet^a|ph\hb_{(2)}=0
 \eel{e:306}
when $\baca^a|ph\hb_{(1)}=\pi_a|ph\hb_{(1)}=0$ and $\pet_a|ph\hb_{(2)}=[Q,
\pet_a]|ph\hb_{(2)}=0$, and the condition that the following sets of hermitian
BRST
doublets
 \be
&&D_{(1)r}\equiv\{\chi^a, i^{\varepsilon_a+1}[Q, \chi^a];
i^{\varepsilon_a}\baca_a, i(-1)^{\varepsilon_a}[Q, \baca_a]=\pi_a\},\nn\\
&&D_{(2)r}\equiv\{\Lambda^a,
i^{\varepsilon_a+1}[Q, \Lambda^a];i^{\varepsilon_a+1}\pet_a, i[Q, \pet_a]\}
\eel{e:307}
each must satisfy a closed algebra with all coefficients to the left.
Furthermore the BRST doublets \r{307} should constitute a  set of generalized
BRST
quartets, \ie $[D_{(1)r}, D_{(2)s}]$ must be an invertible matrix operator.
As in section 3 we require the that the $C$-operators $(\chi^a, \baca_a,
\pet_a,
\Lambda^a)$ essentially commute so that
this condition is satisfied if
\be
&&[\chi^a, [Q, \pet_b]],\;\; [\baca_a, [Q, \Lambda_b]],\;\; [\pet_a, [Q,
\chi^b]],\mbox{ and }[\pi_a, \Lambda^b]\nn\\
&&\mbox{are invertible matrix operators}
\eel{e:308}
This is certainly satisfied if
 \eg $\chi^a$ commutes with antighosts and Lagrange
multipliers and satisfies the  conditions for the minimal sector
and if $\Lambda^a=\pm
i^{\varepsilon_a}\la^a$. However, in addition there are more general solutions
possible here. A nilpotent coBRST operator is here given by an expression like
$^*Q=\chi^a\pet_a+\baca_a\Lambda^a+\ldots$ which requires $\chi^a$
and $\Lambda^a$ to
be involution and independent of $\ca^a$ and $\bapet^a$.

\section{The solutions on inner product spaces.}
\setcounter{equation}{0}
Now even though the above physical states, $|ph\hb_{(1)}$ and
$|ph\hb_{(2)}$, have ghost number zero none of them  belongs to an
inner product space. In fact even if we assume the genuine physical
variables  to span  an inner product space we
still obtain  undefined expressions like \be
 &&_{(l)}\vb ph|ph\hb_{(l)}=0\cdot\infty,\;\;\;l=1,2
\eel{e:401}
On the other hand, as in the minimal sector the bilinear forms
$_{(1)}\vb ph|ph\hb_{(2)}$ are in such a case finite and well
defined. Now according to the results of section 3 the above states
may be used to
 define BRST invariant states on inner product
spaces. Formula \r{6} yields here
 \be
 &&|ph,l\hb=e^{[Q, \psi_l]}|ph\hb_{(l)},\;l=1,2
\eel{e:402}
where $|ph\hb_{(l)}$ are the above physical states as defined
in section 5, and where
$\psi_{(l)}$ are the following hermitian fermionic gauge fixing
operators with ghost
number minus one
 \be
 &&\psi_1=i^{\varepsilon_a+1}\pet_a\Lambda^a
\eel{e:403}
\be
&&\psi_2=i^{\varepsilon_a+1}\baca_a\chi^a
\eel{e:404}
where in turn $\Lambda^a$ and $\chi^a$ are the gauge fixing operators of
section
3 satisfying the
conditions given there.

{}From section 3 we know that the states in \r{402} are formally inner
product solutions for any consistent choice of the gauge fixing
operators $\Lambda^a$ and $\chi^a$ in \r{403} and \r{404}.
Furthermore, since $|ph\hb_{(1,2)}$ need not satisfy the gauge fixing
conditions \r{305} in order to be BRST invariant, one would naively expect the
solutions $|ph, l\hb$ \r{402} to be independent of the gauge fixing
operators  $\Lambda^a$ and $\chi^a$ for $l=1$ and $l=2$
respectively.  In
particular one would expect the
 norms of
$|ph, l\hb$ to be independent of the choice of gauge fixing $\Lambda^a$
and $\chi^a$. Now, this is only true for certain classes of gauge fixings as
will be demonstrated below. However, the gauge independence within such classes
may
be  illustrated by means of some simplifying assumptions. Consider $|ph, 1\hb$
first.
Here we have
\be
&&\vb ph, 1|ph, 1\hb=\,_{(1)}\vb ph|e^{2[Q,
\psi_1]}|ph\hb_{(1)}=\nn\\
&&=\,_{(1)}\vb ph|e^{2i^{\varepsilon_a+1}([Q,
\pet_a]\Lambda^a+[Q, \Lambda^a]\pet_a])}|ph\hb_{(1)} \eel{e:414} If we assume
$\Lambda^a$ to be of the form $\Lambda^a=\halv
(-i)^{\varepsilon_a}X^a_{\;b}\la^b$
where
   $X^a_{\;b}$
is a nonsingular matrix operator with commuting elements and with no dependence
on
the Lagrange multipliers, and which furthermore commutes with $[Q, \pet_a]$,
then we
have
\be
&&\vb ph, 1|ph, 1\hb=\,_{(1)}\vb ph|e^{i[Q,
\pet_a]X^a_{\;b}\la^b+\bapet^bX^a_{\;b}\pet_a]}|ph\hb_{(1)}
\eel{e:415}
The fact that $\bapet^a$, $\pet_a$ have opposite Grassmann parity to
$[Q, \pet_a]$ and
$\la^a$ together with the properties of $|ph\hb_{(1)}$ imply now that
the only dependence on $X^a_{\;b}$ in \r{415} is through the factor:
$\det X^a_{\;b}/(\det X^a_{\;b})=1$. Thus, \r{415} is independent of
$X^a_{\;b}$.  For $|ph, 2\hb$ we have
\be
&&\vb ph, 2|ph, 2\hb=\,_{(2)}\vb ph| e^{2[Q,
\psi_2]}|ph\hb_{(2)}=\nn\\ &&=\,_{(2)}\vb
ph|e^{2(-i)^{\varepsilon_a}(-\pi_a\chi^a+\baca_a
M^a_{\;b}\ca^b)}|ph\hb_{(2)}\propto\int\del(\chi^a)\det
(M^a_{\;b})\:|\phi|^2  \eel{e:416}
where $\phi$ is a matter state which is a solution of the Dirac quantization.
Thus,
here we get under some simplifying assumptions a standard Faddeev-Popov type of
expression which is at least locally independent of $\chi^a$ \cite{Fad}. (The
quantization rules of \cite{Proper} must be used in \r{415} and \r{416}.)

The physical states for the cases 1 and 2, $|ph, 1\hb$ and
$|ph, 2\hb$, should span the same physical state space if they are
obtained from a given original inner product state.  From
section 3 this requires   the complex BRST doublets
$D'_{(1)r}$ and $D'_{(2)r}$ to be algebraically related.  However,
such a relation must be  nonlinear in general which makes the
equivalence  hard to demonstrate. On the other hand, for abelian
gauge theories there exist simple gauge fixing conditions $\chi^a$
for which there are no nonlinear terms in \r{14} and \r{15}. In
this case one may explicitly show that $D'_{(1)r}$ and $D'_{(2)r}$ are
equivalent. (In fact, this is the example given in section 4.) Notice that
when an equivalence is established then we also have an explicit
solution of the Dirac quantization given by \be
&&|ph\hb_{(2)}=e^{-[Q, \psi_2]}e^{[Q, \psi_1]}|ph\hb_{(1)}
\eel{e:4161}

It should be stressed that the physical state space is spanned by \r{402} for
one
specific choice of the gauge fixing operators $\Lambda^a$ and $\chi^a$ in
\r{403}
 and \r{404}. Although there are whole classes of  $\Lambda^a$ and $\chi^a$
which
yield physical states belonging to the same physical inner product space there
 always exist choices which do not.
To demonstrate this consider \eg \r{402} for $\Lambda^a$, $\chi^a$  and
$-\Lambda^a$,
$-\chi^a$, \ie
\be
&&|ph, l\hb=e^{[Q, \psi_l]}|ph\hb_{(l)},\;\;\;|ph,
l\hb'=e^{-[Q, \psi_l]}|ph\hb_{(l)}
 \eel{e:417}
Obviously $|ph, l\hb$ and
$|ph, l\hb'$ do not belong to the same inner product space since
their inner product is undefined, $\vb ph, l|ph, l\hb'=\,_{(l)}\vb
ph|ph\hb_{(l)}=0\cdot\infty$. They are simply spanned by inequivalent
bases which means that the
 corresponding original state spaces are also
spanned by inequivalent bases. This implies that the choice of gauge
fixing is related to the choice of an original inner product space
from which the physical states are projected out. On the other
hand, it seems as if one   always may
impose the condition that the physics of $|ph, l\hb$ and  $|ph,
l\hb'$ should be equivalent \cite{Simple,Proper} although they are
projected from two different state spaces.

Now different choices of gauge fixing lead in general to equivalent state
spaces.
There is \eg always a class of unitary equivalent choices for the gauge
fixing operator
$\psi$: Let $U$ be a BRST invariant unitary operator with ghost number zero,
\ie
\be
&&[Q, U]=0,\;\;\;[N, U]=0,\;\;\;U\dagg U=UU\dagg=\bett
\eel{e:418}
This implies
\be
&&U|ph, l\hb=e^{[Q, \psi'_l]}U|ph\hb_{(l)}
\eel{e:419}
where
\be
&&\psi'_l=U\psi_lU\dagg
\eel{e:420}
Thus, for those $U$ for which
 $U|ph\hb_{(l)}$
satisfies the same conditions as  $|ph\hb_{(l)}$,
$U$  transforms the gauge fixing operators within the
same state space. An important example of such a transformation is
\be &&U=e^{\al [Q, \la^a\baca_a]} \eel{e:421}
where $\al$ is a real parameter. It satisfies \r{418} and since
\be
&&[Q, \la^a\baca_a]=i\la^a\pi_a-i(-1)^{\varepsilon_a}\bapet^a\baca_a
\eel{e:422}
we have
\be
&&U|ph\hb_1=|ph\hb_1
\eel{e:423}
when $|ph\hb_1$ satisfies \r{303}, and
\be
&&U|ph, \la, \pi\hb_2=|ph, e^{-\al}\la, e^{\al}\pi\hb_2
\eel{e:424}
when $|ph\hb_2$ satisfies \r{304}. (The Lagrange multiplier
dependence in $|ph\hb_2$
represents unphysical gauge degrees of freedom. It may be fixed by the
condition
\r{305}.) For the gauge fixing fermions \r{403} and \r{404} we simply get a
scaling
\be
&&\psi'_1=U\psi_1U\dagg=e^{-\al}\psi_1,\;\;\;\psi'_2=U\psi_2U\dagg=e^{\al}\psi_2
\eel{e:425}
provided $\chi^a$ does not involve the Lagrange multipliers in other
combinations
than $\la^a\pi_b$, and provided $\Lambda^a$ is linear in $\la^a$.
Thus, we may always
scale the exponents in \r{402} without affecting the norms and physical
contents of
the states \cite{Simple,Proper}.

 \section{The general reducible case} \setcounter{equation}{0}
The above results for the  irreducible case may also be
extended to the general reducible case. Let us consider a general
L-stage reducible gauge theory. Here the basic gauge generators
$\theta_{a_0}$, $a_0=1,\ldots,m_0$, satisfy not only the involution
relations but also the additional conditions \cite{BF3}
\be
&&\theta_{a_0}Z^{a_0}_{1a_1}=0,\;\;\;a_1=1,\ldots,m_1\nn\\
&&Z^{a_{s-2}}_{s-1\:a_{s-1}}Z^{a_{s-1}}_{sa_s}=0,\;\;\;a_s=1,\ldots,m_s,
\;\;s=2,\ldots,L
\eel{e:501}
where
\be
&&\mbox{rank}\,Z^{a_{s-1}}_{sa_s}=\ga_s(L),\;\;\;
\ga_s(L)\equiv\sum^L_{s'=s}m_{s'}(-1)^{(s'-s)}
\eel{e:502}
These relations imply that the $m_0$ constraint variables
$\theta_{a_0}$ are dependent and only represent $\ga_0(L)<m_0$
irreducible constraints. An invariant BRST charge requires us to
introduce $m_0$ ghost variables to $\theta_{a_0}$ which is too
many. On the other hand these extra ghosts may be compensated by
the introduction of ghosts for ghosts. The resulting nilpotent BRST
charge operator in the minimal sector involves then the
following total set of ghosts
and their conjugate momenta  \cite{BF3}:
\be
&&\ca_s^{a_s},\;\pet_s^{a_s},\;\;s=0,\ldots,L;\;a_s=1,\ldots,m_s
\eel{e:503}
which satisfy
\be
&&[\ca^{a_s}_s, \pet_{s'a_{s'}}]=i\del_{ss'}\del^{a_s}_{a_{s'}}
\eel{e:504}
The BFV-BRST charge is of the form
\be
&&\Om=\ca^{a_0}_0\theta_{a_0}+\sum_{s=0}^{L-1}\ca_{s+1}^{a_{s+1}}
Z_{a_{s+1}}^{{a_s}} \pet_{sa_{s}} +\mbox{higher order terms in the
ghosts}
\eel{e:505}
The required ghost numbers and Grassmann parities are
\be
&&gh(\ca^{a_s}_s)=s+1,\;\;\;gh(\pet_{sa_{s}})=-(s+1)\nn\\
&&\ve(\ca^{a_s}_s)=\ve_{a_s}+s+1=\ve(\pet_{sa_{s}})
\eel{e:506}
where $\ve_{a_s}$ is defined by
\be
&&\ve_{a_0}\equiv\ve(\theta_{a_0})\nn\\
&&\ve_{a_s}\equiv\ve(Z_{a_{s+1}}^{{a_s}})-\ve_{a_{s-1}},\;\;\;s\geq 1
\eel{e:507}
$\theta_{a_0}$ and $\ca^{a_s}_s$ are assumed to be hermitian which implies
$(\pet_{sa_{s}})\dagg=(-1)^{\ve_{a_s}}\pet_{sa_{s}}$ from \r{504}.

We try now to fix the ghost dependence of the physical states.
As in the irreducible
case there are only two sets of conditions which may be possible to
impose in general.
They are
\be
(1)&&\;\;\;\;\;\ca_s^{a_s}|ph\hb_{(1)}=0,\;\;\;s=0,
\ldots,L;\;\;a_s=1,\ldots,m_s
\eel{e:508}
\be
(2)&&\;\;\;\;\;\pet_{sa_s}|ph\hb_{(2)}=0,\;\;\;s=0,
\ldots,L;\;\;a_s=1,\ldots,m_s
\eel{e:509}
Consistency requires
\be
(1)&&\;\;\;\;\;[\Om,
\ca_s^{a_s}]|ph\hb_{(1)}=0,\;\;\;s=0,\ldots,L;\;\;a_s=1,\ldots,m_s
\eel{e:510} \be
(2)&&\;\;\;\;\;[\Om,
\pet_{sa_s}]|ph\hb_{(2)}=0,\;\;\;s=0,\ldots,L;\;\;a_s=1,\ldots,m_s
\eel{e:511}
Eq. \r{510} is automatically satisfied (use a $\pet\ca$-ordered
$\Om$) while \r{511} only contains the following new nontrivial condition
\be
&&[\Om, \pet_{0a_0}]|ph\hb_{(2)}\equiv(\theta_{a_0}+\ldots
)|ph\hb_{(2)}=  0
\eel{e:512}
in addition to \r{509} (use a $\ca\pet$-ordered
$\Om$)

Let us now look for consistent sets of BRST doublets. In case (2)
we have
\be
&&[\Om,
\pet_{saQs}]=iZ_{a_s}^{a_{s-1}}\pet_{s-1\,a_{s-1}}+(\;)\pet^2
\eel{e:513}
which implies that not all $\pet_{sa_s}$ are C-operators. In fact,
since
\be
&&\mbox{rank}Z_{a_s}^{a_{s-1}}=\ga_s(L)
\eel{e:514}
only $\ga_s(L)$ $\pet_{sa_s}$-operators are C-operators. This is
less than $m_s$ since
\be
&&\ga_s(L)+\ga_{s+1}(L)=m_s,\;\;\;\ga_L(L)=m_L
\eel{e:515}
from \r{502}. We define these C-operators to be
\be
&&P_{sa_s}\equiv R_{sa_s}^{b_s}\pet_{sb_s}
\eel{e:516}
where $R_{sa_s}^{b_s}$ satisfies
\be
&&\mbox{rank}
R_{sa_s}^{b_s}=\ga_s(L),\;\;\;
\mbox{rank}R_{sa_s}^{b_s}Z_{b_s}^{a_{s-1}}=\ga_s(L)\;\;\;
\ve(R_{sa_s}^{b_s})=\ve_{a_s}+\ve_{b_s}
\eel{e:517}
We also require $R_{sa_s}^{b_s}$ to be such that $P_{sa_s}$ is hermitian.
Now, due to \r{515} there are
$\ga_{s+1}(L)$ $\pet_{sa_s}$-operators which are B-operators and from \r{513}
they are
given by $Z_{b_s}^{a_{s-1}}(+\cdots)\pet_{sb_s}$.
The total set of doublets in case (2) are then
$\{P_{sa_s}; [\Om, P_{sa_s}]\}
$.
Notice that they are equivalent to \r{509} and \r{512}. The
counting agrees since the total number of constraint operators are
\be
&&2\sum_{s=0}^L\ga_s(L)=\ga_0(L)+\sum_{s=0}^Lm_s
\eel{e:519}
(For each $s$ there are $\ga_s(L)$ doublets, and there are $m_s$
$\pet_{sa_s}$-operators for each $s$ plus $\ga_0(L)$ irreducible
original constraints.)

For the ghost part (case (1)) we introduce hermitian gauge fixing operators
$\chi_s^{a_s}$ satisfying
\be
&&[\Om, \chi_s^{a_s}]=
iK_{sb_s}^{a_s}\ca_s^{b_s}\nn\\
&&gh(\chi_s^{a_s})=s,\;\;\;\mbox{rank} K_{sb_s}^{a_s}=\ga_s(L)
\eel{e:520}
Notice that $K_{0b_0}^{a_0}$ can only have rank $\ga_0(L)$ since
$\theta_{a_0}$ only involves $\ga_0(L)$ independent components.
Thus, $\chi_0^{a_0}$ involves only $\ga_0(L)$ independent components and are as
in the irreducible case also gauge fixing conditions to
$[\Om, P_{0a_0}]$. Since
\be
&&[\Om, \ca_s^{a_s}]=\ca_{s+1}^{a_{s+1}}Z_{a_{s+1}}^{a_s}
\eel{e:521}
it follows that we may choose
\be
&&\chi_s^{a_s}\equiv\omega_{sa_{s-1}}^{a_s}\ca_{s-1}^{a_{s-1}},\;\;\;s\geq 1
\eel{e:522}
where
\be
&&\mbox{rank}\,\omega_{sa_{s-1}}^{a_s}=\ga_s(L),\;\;\;
\mbox{rank}\,Z_{b_{s}}^{a_{s-1}}\omega_{sa_{s-1}}^{a_s}=\ga_s(L)
\eel{e:523}
which implies that rank$ K_{sb_s}^{a_s}=\ga_s(L)$ in \r{520}.
Obviously $(\chi_s^{a_s},
K_{sb_s}^{a_s}\ca_s^{b_s})$ constitute $\ga_s(L)$ BRST doublets for
each $s$. These variables are equivalent to $\ca_s^{a_s}$ for all $s$ plus
$\chi_0^{a_0}$. (The counting agrees due to the equality \r{519}.)
We conclude that in order to project out BRST singlets the ghost
fixing \r{508} and \r{509} must be accompanied by the additional conditions
\be
&&(1)\;\;\;\chi_0^{a_0}|ph\hb_{(1)}=0,\;\;a_0=1,\ldots,m_0\nn\\
&&(2)\;\;\;[\Om, \pet_{0 a_0}]|ph\hb_{(2)}=0,\;\;a_0=1,\ldots,m_0
\eel{e:524}

We have ,thus, found the following two sets of BRST doublets
\be
&&D_{(1)r}=\{\chi_s^{a_s}, [\Om, \chi_s^{a_s}],\;a_s=1,\ldots,m_s;
\:s=0,\dots,L\}\nn\\
&&D_{(2)r}=\{P_{s a_s}, [\Om, P_{s a_s}],\;\;a_s=1,\ldots,m_s;\:s=0,\dots,L\}
\eel{e:525}
Each of them are required to be in involution, and with appropriate
factors of i they may be chosen to be hermitian. We require now that
they are dual sets of doublets so that they together form BRST
quartets. In other words we require the matrix operator
$[D_{(1)r}, D_{(2)s}]$ to be invertible. As before we demand that
$[\chi_s^{a_s}, P_{s' a_{s'}}]$ essentially vanish
so that it is sufficient to require
\be
&&\mbox{rank}[P_{s a_{s}}, [\Om, \chi_{s'}^{b_{s'}}]]=\ga_s(L)\nn\\
&&\mbox{rank}[\chi_s^{a_s}, [\Om, P_{s' b_{s'}}]]=\ga_s(L)
\eel{e:527}
These conditions are satisfied by the conditions which
we have already imposed. Notice
that
 \be
&&\mbox{rank}[\chi_0^{a_0}, [\Om, P_{0 b_{0}}]]=\ga_0(L)
\eel{e:528}
follows from rank$K_{0b_0}^{a_0}=\ga_0(L)$ in \r{520} due to the Jacobi
identities.
 That $[\chi_s^{a_s}, P_{s' a_{s'}}]$ essentially vanishes requires the
following connections between the matrix operators  $\omega_{sb_{s-1}}^{a_s}$
and
$R_{s\,b_{s}}^{aQ{s}}$: \be
&&\omega_{sb_{s-1}}^{a_s}R_{s-1\,a_{s-1}}^{b_{s-1}}=0
\eel{e:526}\\

We turn now to the nonminimal sector. In \cite{BF3} it was shown that the
correct
form of the extended BRST charge is given by
\be
&&Q=\Om+\sum_{s=0}^L\pi_{sa_s}\bapet^{a_s}_s+\sum_{s'=1}^L\sum_{s=s'}^L\pi_{s
a_s}^{s'}\bapet_s^{s' a_s}
 \eel{e:531}
where the new variables have the properties ($\pi_{sa_s}$ and
$\bapet^{a_s}_s$ are chosen to be hermitian)
\be
&&[\la_s^{b_{s}}, \pi_{r
a_r}]=i\del_{sr}\del_{a_r}^{b_{s}},
\;\;\;[\baca_{sa_s}, \bapet_{r
}^{b_{r}}]=i\del_{sr}\del^{b_{r}}_{a_s}\nn\\
&& gh(\pi_{s
a_s})=- gh(\la_s^{a_{s}}) =-s,
\nn\\
&& gh(\baca_{sa_s})=-
gh(\bapet_{s }^{a_{s}})=-(s+1)\nn\\
&&\ve(\pi_{s
a_s})=\ve(\la_s^{a_{s}}) =\ve_{a_s}+s,
\nn\\
&& \ve(\baca_{sa_s})=
\ve(\bapet_{s }^{a_{s}})=\ve_{a_s}+s+1\nn\\
&&a_s=1,\ldots,m_s,\;\;\;s=0,\ldots,L
\eel{e:530}
and
($\pi_{s
a_s}^{s'}$ and $\bapet_s^{s' a_s}$ are chosen to be hermitian)
\be
&&[\la_s^{s'b_{s}}, \pi_{r
a_r}^{r'}]=i\del^{s'r'}\del_{sr}\del_{a_r}^{b_{s}},
\;\;\;[\baca_{sa_s}^{s'}, \bapet_{r
}^{r'b_{r}}]=i\del^{s'r'}\del_{sr}\del^{b_{r}}_{a_s}\nn\\
&& gh(\pi_{s
a_s}^{s})=- gh(\la_s^{s'a_{s}}) =-(s'-s),
\nn\\
&& gh(\baca_{sa_s}^{s'})=-
gh(\bapet_{s }^{s'a_{s}})=-(s-s'+1)\nn\\
&&\ve(\pi_{s
a_s}^{s'})=\ve(\la_s^{s'a_{s}}) =\ve_{a_s}+s-s',
\nn\\
&& \ve(\baca_{sa_s}^{s'})=
\ve(\bapet_{s }^{s'a_{s}})=\ve_{a_s}+s-s'\nn\\
&&a_s=1,\ldots,m_s,\;\;\;s=s',\ldots,L,\;\;\;s'=1,\ldots,L
\eel{e:532}

All these new variables  are unphysical. They may be grouped into
additional BRST doublets, $(\baca, \pi)$ and $(\la,\bapet)$, which
commute with the ones in the minimal sector. The question is now how to
combine these doublets with the original ones in cases (1) and (2)
so that the physical states will have ghost number zero. Below we
give a simple algoritm how to split the doublets appropriately.

Guided by the irreducible case we  first add $(\baca_{sa_s}, \pi_{sa_s})$ to
$D_{(1)r}$ and
$(\la_s^{a_s}, \bapet_s^{a_s})$ to $D_{(2)r}$,  where $D_{(1,2)r}$ are defined
in
\r{525}. We apply then the general formula \r{6} in section 3. We find then the
hermitian gauge fixing operators ($P_{sa_s}$ and
$\chi_s^{a_s}$ are here chosen such that $\psi_{1,2}$ are
hermitian) \be
&&\psi_1=\sum_{s=0}^LP_{sa_s}\la_s^{a_s}\nn\\
&&\psi_2=\sum_{s=0}^L\baca_{sa_s}\chi_s^{a_s}
\eel{e:533}
Now since $\chi_s$ and $P_s$ each only represents $\ga_s(L)$ independent
variables, $\ga_{s+1}(L)$ of the $\la_s$ and $\baca_s$ variables are
not involved in \r{533}.
In order to introduce also them into the gauge fixing function we need
auxiliary variables. The additional terms must have the form
\be
&&\triangle_1\psi_1=\sum_{s=0}^{L-1}P^1_{sa_s}\la_s^{a_s},\;\;\;
gh(P^1_{sa_s})=-(s+1)\nn\\
&&\triangle_1\psi_2=\sum_{s=0}^{L-1}\baca_{sa_s}\chi^{1a_s}_s,\;\;\;
gh(\chi^{1a_s}_s)=s
\eel{e:534}
where $\chi^{1a_s}_s$ and $P^1_{sa_s}$ each represents
$\ga_{s+1}(L)$ new degrees of freedom. Since we have only covariant
variables at our disposal we define (cf. \cite{BF3})
\be
&&\chi^{1a_{s-1}}_{s-1}\equiv
\bar{\omega}^{1a_{s-1}}_{s-1\:a_s}\la^{1 a_s}_s
,\;\;\;\mbox{rank}\,\bar{\omega}^{1a_{s-1}}_{s-1\:a_s}=\ga_s(L)\nn\\
&&P^1_{s-1\:a_{s-1}}\equiv \baca^{1}_{sa_{s}}\sigma^{1a_{s}}_{sa_{s-1}}
,\;\;\;\mbox{rank}\,{\sigma^{1a_{s}}_{sa_{s-1}}}=\ga_s(L) \eel{e:535}
However, now we have introduced too many
$\la^{1 a_s}_s$ and $\baca^{1}_{sa_{s}}$
variables. We must therefore add
\be
&&\triangle_2\psi_1=\sum_{s=1}^{L-1}\baca^1_{sa_s}
\chi^{2a_s}_s,\;\;\;gh(\chi^{2a_s}_s
)=s-1\nn\\
&&\triangle_2\psi_2=\sum_{s=1}^{L-1}P^2_{sa_s}
\la_s^{1\,a_s},\;\;\;gh(P^2_{sa_s})
=-s
\eel{e:536}
where
\be
&&\chi^{2a_{s-1}}_{s-1}\equiv
\bar{\omega}^{2a_{s-1}}_{s-1\:a_s}\la^{2 a_s}_s
,\;\;\;\mbox{rank}\,\bar{\omega}^{2\,a_{s-1}}_{s-1\:a_s}=\ga_s(L)\nn\\
&&P^2_{s-1\:a_{s-1}}\equiv \baca^{2}_{sa_{s}}\sigma^{2a_{s}}_{sa_{s-1}}
,\;\;\;\mbox{rank}\,{\sigma^{2\,a_{s}}_{sa_{s-1}}}=\ga_s(L)
\eel{e:537}
But with \r{536} we have introduced too many $\la^{2 a_s}_s$
and $\baca^{2}_{sa_{s}}$
variables which forces us to add further terms. This procedure
eventually stops and
we arrive at the final formulas
\be
&\psi_1=&\sum_{s=0}^LP_{sa_s}\la_s^{a_s}+
\sum_{s=0}^{L-1}P^1_{sa_s}\la_s^{a_s}+
\sum_{s=1}^{L-1}\baca^1_{sa_s}\chi^{2a_s}_s+\sum_{s=2}^{L-1}
P^3_{sa_s}\la_s^{2a_s}+\nn\\
&&+\sum_{s=3}^{L-1}\baca^3_{sa_s}\chi^{4a_s}_s+\ldots=
\sum_{s=0}^LP_{sa_s}\la_s^{a_s}+\nn\\
&&+
\sum_{s'=0}^{[(L-1)/2]}\sum_{s=2s'+1}^{L}\left(\baca^{2s'+1}_{sa_s}
\sigma^{2s'+1\:a_s}_{sa_{s-1}}
\la^{2s'\,a_{s-1}}_{s-1}\right)+\nn\\
&&+\sum_{s'=1}^{[L/2]}\sum_{s=2s'}^{L}\left(\baca^{2s'-1}_{s-1\:a_{s-1}}
\bar{\omega}^{2s'\,a_{s-1}}_{s-1\:a_{s}}\la^{2s'\,a_{s}}_{s}
\right)
\eel{e:538}
\be
&\psi_2=&\sum_{s=0}^L\baca_{sa_s}\chi_s^{a_s}+\sum_{s=0}^{L-1}
\baca_{sa_s}\chi^{1a_s}_s+\sum_{s=1}^{L-1}P^2_{sa_s}\la_s^{1a_s}+
\sum_{s=2}^{L-1}\baca^2_{sa_s}\chi^{3a_s}_s+
\nn\\
&&+\sum_{s=3}^{L-1}P^4_{sa_s}\la_s^{3a_s}+\dots=\sum_{s=0}^L
\baca_{sa_s}\chi_s^{a_s}+\nn\\
&&+
\sum_{s'=1}^{[L/2]}\sum_{s=2s'}^{L}\left(\baca^{2s'}_{sa_s}
\sigma^{2s'\,a_s}_{sa_{s-1}}
\la^{2s'-1\:a_{s-1}}_{s-1}\right)+\nn\\
&&+\sum_{s'=0}^{[(L-1)/2]}\sum_{s=2s'+1}^{L}
\left(\baca^{2s'}_{s-1\:a_{s-1}}
\bar{\omega}^{2s'+1\,a_{s-1}}_{s-1\:a_{s}}\la^{2s'+1\:a_{s}}_{s}
\right)
\eel{e:539}
where
\be
&&\chi^{s'\,a_{s-1}}_{s-1}\equiv
\bar{\omega}^{s'\,a_{s-1}}_{s-1\:a_s}\la^{s'\, a_s}_s
,\;\;\;\mbox{rank}\,\bar{\omega}^{2a_{s-1}}_{s-1\:a_s}=\ga_s(L)\nn\\
&&P^{s'}_{s-1\:a_{s-1}}\equiv \baca^{s'}_{sa_{s}}
\sigma^{s'\,a_{s}}_{sa_{s-1}}
,\;\;\;\mbox{rank}\,{\sigma^{1a_{s}}_{sa_{s-1}}}=\ga_s(L)\nn\\
&&\la^{0\:a_s}_s\equiv\la^{a_s}_s,\;\;\;
\baca^{0}_{sa_{s}}\equiv\baca_{sa_{s}}
\eel{e:540}

Obviously the auxiliary variables in \r{532} are exactly what was needed to
get the counting right. In fact, they could have been derived by
this argument.

{}From the expressions \r{538} we notice that $\psi_1$ involves the
additional C-operators $(\la^{2s'},\baca^{2s'-1})$,
$s'=1,\ldots,[L/2]$, while
$\psi_2$ involves $(\la^{2s'-1},\baca^{2s'})$, $s'=1,\ldots, [L/2]$.
The final set of effective
BRST doublets for cases (1) and (2) are then
\be
&(1)&\;\;\;\{(\chi_s^{a_s}, [\Om, \chi_s^{a_s}]); (\baca^{2s'}_{sa_s},
\pi^{2s'}_{sa_s}); (\bapet^{2s'-1\:a_s}_s, \la^{2s'-1\:a_s}_s)\}\nn\\
&(2)&\;\;\;\{(P_{sa_s}, [\Om, P_{sa_s}]); (\baca^{2s'-1}_{sa_s},
\pi^{2s'-1}_{sa_s}); (\bapet^{2s'\,a_s}_s, \la^{2s'\,a_s}_s)\}
\eel{e:541}
This implies that the singlet states $|ph\hb_{(1,2)}$ are determined by the
conditions
\be
&(1)&\;\;\;\chi_0^{a_0}|ph\hb_{(1)}=\ca^{a_s}_s
|ph\hb_{(1)}=\baca_{sa_s}|ph\hb_{(1)}
=\pi_{sa_s}|ph\hb_{(1)}=\nn\\
&&=\baca^{2s'}_{sa_s}|ph\hb_{(1)}=\pi^{2s'}_{sa_s}|ph\hb_{(1)}=
\bapet^{2s'-1\:a_s}_s|ph\hb_{(1)}=\la^{2s'-1\,a_s}_s|ph\hb_{(1)}=0
\eel{e:542}
\be
&(2)&\;\;\;[\Om,
P_{0a_0}]|ph\hb_{(2)}=\pet_{sa_s}|ph\hb_{(2)}=\bapet^{a_s}_s
|ph\hb_{(2)}=\la^{a_s}_s
|ph\hb_{(2)}=\nn\\
&&=\bapet^{2s'\,a_s}_s
|ph\hb_{(2)}=\la^{2s'\,a_s}_s
|ph\hb_{(2)}=\baca^{2s'-1}_{sa_s}
|ph\hb_{(2)} =\pi^{2s'-1}_{sa_s}|ph\hb_{(2)}=0
 \eel{e:543}
One may easily check that these conditions imply that
$|ph\hb_{(1)}$ and $|ph\hb_{(2)}$
have ghost number zero. The physical states on an inner product
space are then given by
\be
&&|ph, l\hb=e^{[Q, \psi_l]}|ph\hb_{(l)},\;\;\;l=1,2
\eel{e:544}

The reducible case on inner product spaces has also been treated in \cite{FS}
and
\cite{FHP}. In \cite{FS} the linear case
(perturbative unitarity) is treated and in
\cite{FHP}  some traces of case (2) is given within the path integral
formulation,  the
conditions \r{543} are \eg imposed as boundary conditions (cf. \cite{Path}).

 \section{Conclusions} \setcounter{equation}{0}
We have derived general formal solutions of BRST quantizations on inner product
spaces by means of a generalized quartet mechanism. For this our starting point
was
that the physical states, which constitute a representation of the BRST
cohomology,
are determined by conditions of the form
\be
&&D_i|ph\hb=0
\eel{e:601}
where $\{D_i\}$ is a complete set of BRST doublets
in involution, or by the conditions
\be
&&D_i^{\dag}|ph\hb=0
\eel{e:602}
depending on the choice of basis for the original state space from which the
projection to the physical states is performed.    $D_i\dagg$ are required to
be
algebraically independent of $D_i$ and to form together with $D_i$  a complete
set
of generalized BRST quartets. The conditions \r{601} and \r{602}
should always be the
consequences of a more invariant formulation like the coBRST one which yields
\be
&&Q|ph\hb=\,^*Q|ph\hb=0
\eel{e:603}
where $Q$ and $^*Q$ are the nilpotent BRST and coBRST operators.

 From \r{601} and \r{602} we have found that the physical
states may be written as follows
 \be
&&|ph\hb=e^{[Q,
\psi]} |ph\hb_0
\eel{e:604}
where $\psi$ is a  hermitian fermionic gauge fixing operator, and $|ph\hb_0$
BRST
invariant states determined by a {\em hermitian} set of BRST doublets in
involution.
What concerns the unphysical degrees of freedom  $|ph\hb_0$
does not in general belong
to an inner product space although $|ph\hb$ does. Since the
BRST quartets may also be
split into two sets of hermitian BRST doublets there are two
choices for $|ph\hb_0$
and the corresponding $\psi$.

The expression \r{604} might seem  to be a way  to just rewrite
the conditions \r{601}
or \r{602} in terms of similar conditions for $|ph\hb_0$.
This is right. However, the
point is that the conditions on $|ph\hb_0$ are much simpler to solve,
not the least
since $|ph\hb_0$ is not restricted to be an inner product state. This we
have demonstrated by a detailed analysis of general gauge theories
given within the BFV
formulation. We have analyzed both  irreducible and
reducible  gauge theories of arbitrary rank and found that there always exist
one
set of solutions,
 $|ph\hb_0$, which
are trivial BRST invariant states  which only depends on the matter
variables , and
another set,  $|ph\hb_0$, which are solutions of a Dirac quantization.
  These solutions generalize the solutions for Lie group theories given in
\cite{Simple,Gauge} but there obtained by means of a bigrading.

There are several aspects of this approach which remains to elaborate. The
connections with the coBRST formulation should \eg be further clarified, and
the freedom in the choice of gauge fixing fermion $\psi$ should be determined.
Notice that we have only given the necessary ingredients of $\psi$
for a given set  of
$|ph\hb_0$ solutions. It is clear that
$\psi$ may be chosen in many more ways.  It remains to give a
precise definition of the time evolution
in terms of a nontrivial Hamiltonian. Notice that the Hamiltonian has not
entered
our treatment so far. The generalization to infinite degrees of
freedom should also involve some technicalities to be clarified.

  \end{document}